\documentclass{article}

\usepackage{mathpple}
\usepackage{graphicx}
\usepackage{url}
\usepackage{paralist}

\usepackage{geometry}
\begin{document}

\noindent
\textbf{Preprint of:}\\
Vincent L. Y. Loke, Timo A. Nieminen, Agata M. Bra\'{n}czyk,\\
Norman R. Heckenberg and Halina Rubinsztein-Dunlop\\
``Modelling optical micro-machines''\\
pp. 163--166 in\\
Nikolai Voshchinnikov (ed.)\\
\textit{9th International Conference on Electromagnetic and Light S
cattering by Non-Spherical Particles: Theory, Measurements, and Applications}\\
(St. Petersburg State University, St. Petersburg, 2006).\\
Online at: \url{http://www.astro.spbu.ru/ELS9/}

\hrulefill

\begin{center}

\Large
\textbf{Modelling optical micro-machines}

\vspace{2mm}

\large
Vincent L. Y. Loke, Timo A. Nieminen, Agata M. Bra\'{n}czyk,\\ Norman R. Heckenberg and Halina Rubinsztein-Dunlop

\vspace{1mm}

\normalsize
\textit{Centre for Biophotonics and Laser Science, School of
Physical Sciences,\\ The University of Queensland, QLD 4072,
Australia}

\end{center}

\begin{abstract}
A strongly focused laser beam can be used to trap, manipulate and
exert torque on a microparticle. The torque is the result of
transfer of angular momentum by scattering of the laser beam.
The laser could be used to drive a rotor, impeller, cog wheel
or some other microdevice of a few microns in size, perhaps
fabricated from a birefringent material. We review our methods
of computationally simulating the torque and force imparted by
a laser beam. We introduce a method of hybridizing the T-matrix
with the Finite Difference Frequency Domain (FDFD) method to
allow the modelling of materials that are anisotropic and
inhomogeneous, and structures that have complex shapes. The
high degree of symmetry of a microrotor, such as discrete or
continuous rotational symmetry, can be exploited to reduce
computational time and memory requirements by orders of
magnitude. This is achieved by performing calculations for
only a given segment or plane that is repeated across the
whole structure. This can be demonstrated by modelling the
optical trapping and rotation of a cube.
\end{abstract}

\section{Introduction}

The T-matrix method \cite{mish1991} is commonly used to calculate properties of light scattering from axisymmetric mesoscale ($\frac{1}{2}\lambda - 5\lambda$) particles that are homogeneous and isotropic \cite{niem2003} \cite{niem2004}. Using the T-matrix, the optical force and torque imparted on the particle by the incident beam can be calculated \cite{niem2004}. The T-matrix is independent of the incident field and only dependent of the properties (size, shape, orientation, permittivity) of the particle. If the incident fields change, the T-matrix need not be recalculated. 

We extend this method to model particles that are inhomogenous, anisotropic and have complex geometrical shapes by combining the T-matrix method with the Finite Difference Frequency Domain (FDFD) method. In the FDFD method, we discretize the computational region into a grid with sufficiently small grid size. The inclusion of FDFD equations in the algorithm is computationally intensive. To optimize computational time and memory usage, we consider the rotational symmetry of the system. If the particle is rotationally symmetric about an axis, the system  could be reduced to a 2D problem (figures 1a and 1b). By choosing a cylindrical coordinate system, the section of interest can be treated in 2D rectangular $(r,z)$ coordinates which leads to compatibility with the FDFD cell (figure 1c). If the particle has \textit{n}th-order discrete rotational symmetry, typical of a microrotor, savings in computational time and memory could still be achieved by performing calculations for only one repeated segment. For example, we modelling the optical trapping of a cube, exploiting the 4th order rotational symmetry and $xy$-plane mirror symmetry, to reduce the time required to calculate the T-matrix from 30 hours to 20 minutes. 

\section{FDFD equations}

\begin{figure}[!htbp]
\begin{center}
a)\includegraphics[height=6cm]{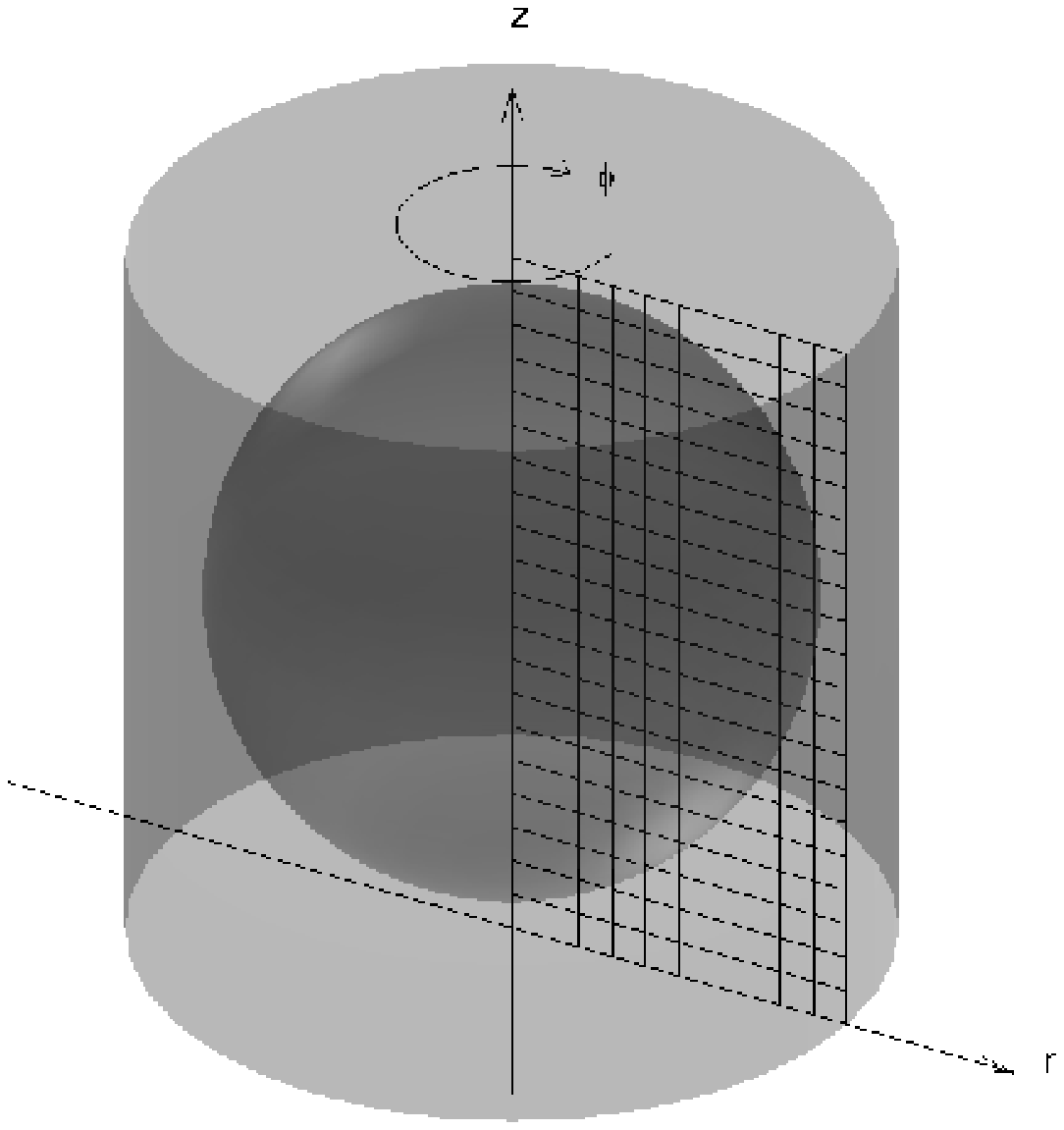}
b)\includegraphics[height=6cm]{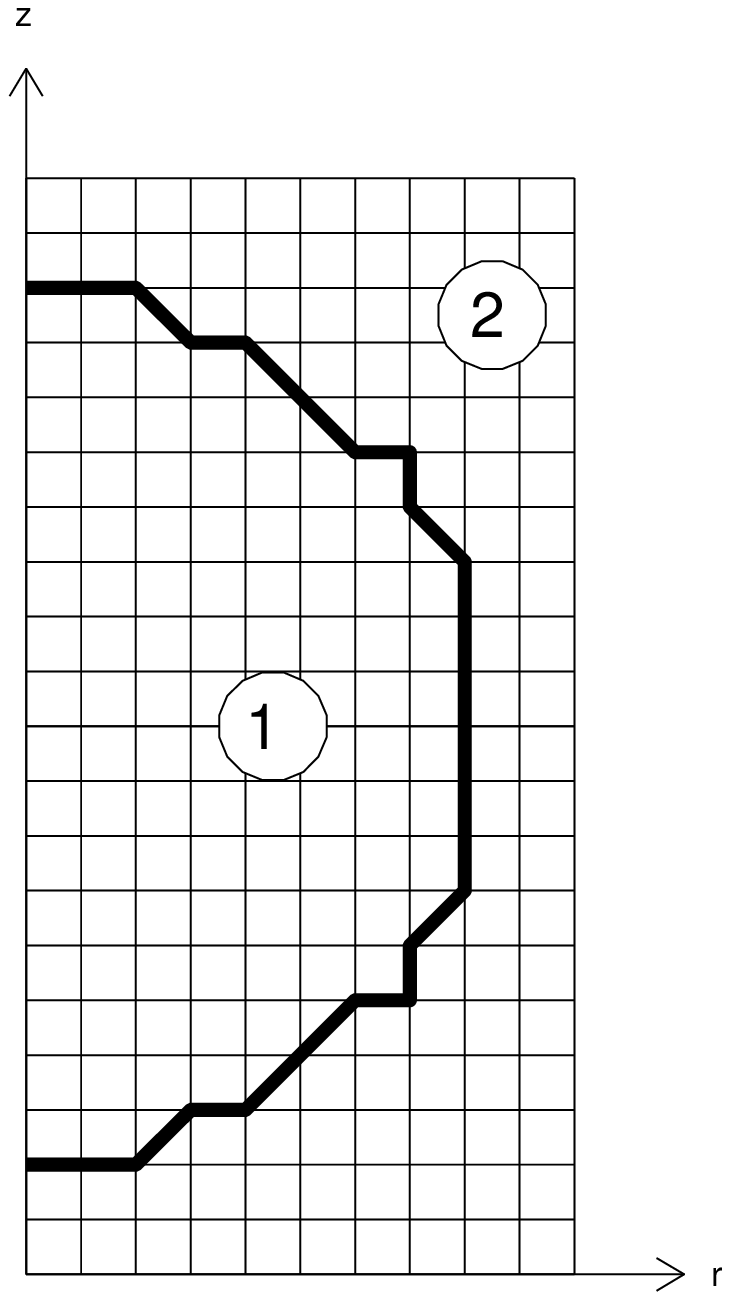}
c)\includegraphics[height=6cm]{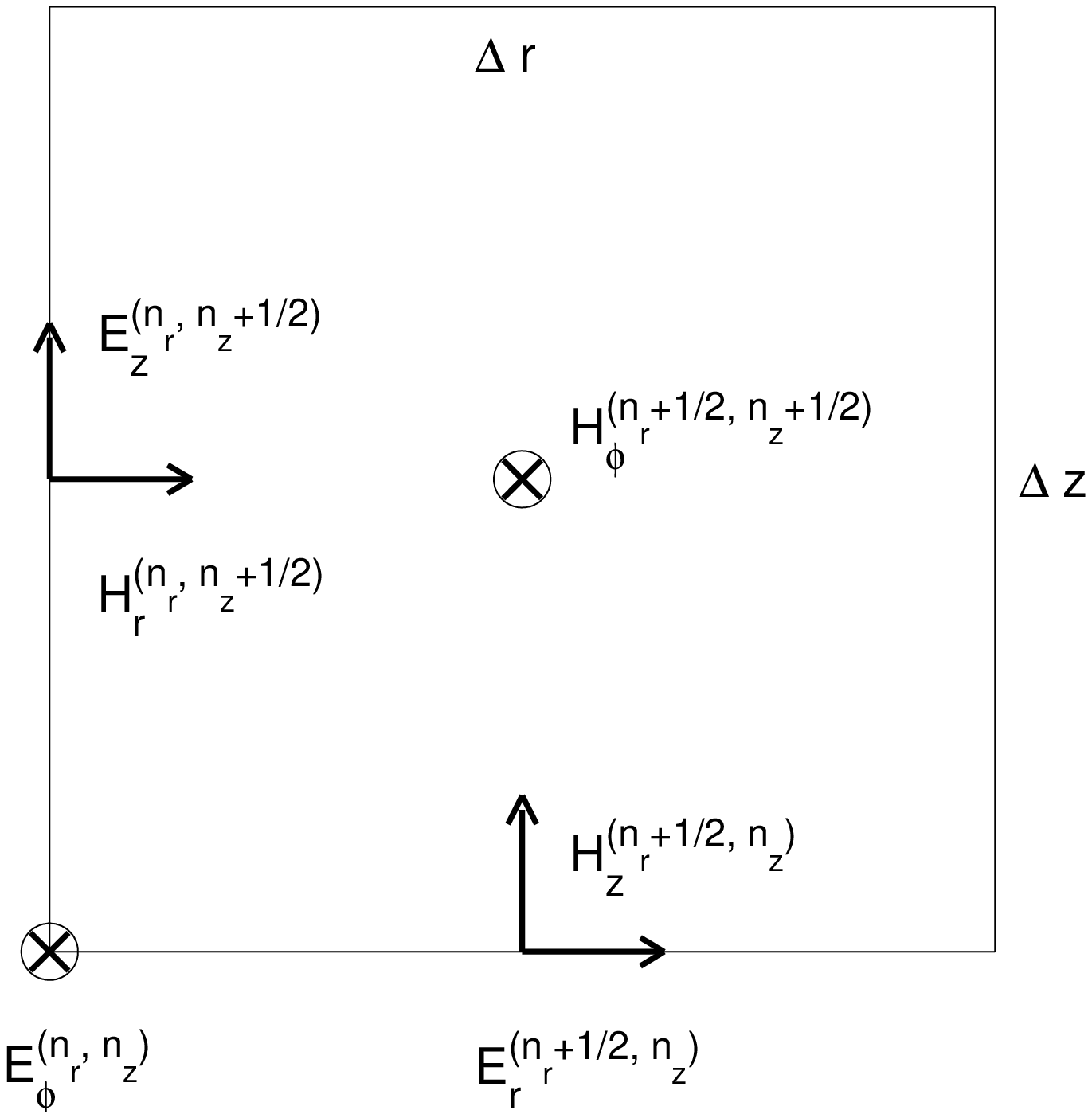}
\caption{a) Cylindrical coordinate system. Spheroidal particle
enclosed in a cylindrical volume. b) Rectangular computational
grid with regions (1) and (2), inside and outside the particle
respectively. c) FDFD (Yee) cell.}
\end{center}
\end{figure}

The rotationally symmetric FDFD equations were derived by expanding Maxwell curl and divergence equations  \cite{jack1998} in cylindrical coordinates
\begin{equation}
\nabla\times \vec{A}=
\left(\frac{1}{r}\frac{\partial \vec{A}_z}{\partial\phi} - \frac{\partial \vec{A}_\phi}{\partial z}\right)\hat{r} +
\left(\frac{\partial \vec{A}_r}{\partial z} - \frac{\partial \vec{A}_z}{\partial r}\right)\hat{\phi} -
\frac{1}{r}\left(\frac{\partial \left(r\vec{A}_\phi\right)}{\partial r} - \frac{\partial \vec{A}_r}{\partial \phi}\right)\hat{z},
\end{equation}
\begin{equation}
\nabla\cdot\vec{A}=\frac{1}{r}\frac{\partial}{\partial r}(r\vec{A_r})+\frac{1}{r}\frac{\partial \vec{A}_{\phi}}{\partial \phi}+\frac{\partial \vec{A}_z}{\partial z},
\end{equation}
where $\vec{A}$ represents the electric field, $\vec{E}$, or the magnetic field, $\vec{H}$. The evolution of both fields can be expressed as $\partial\vec{A}/\partial t = -\mathrm{i}\omega\vec{A}$. It is sufficient, as will be seen below, to consider a field with azimuthal variation $\exp(\mathrm{i}m\phi)$; the variation of the field with respect to $\phi$ would be $\partial\vec{A}/\partial\phi = \mathrm{i}m\vec{A}$.
Substituting the time evolution and $\phi$ harmonic equations into Maxwell's equations we obtain 6 curl equations and 2 divergence equations for electric and magnetic fields. As an example, the curl equation for $\vec{E}_r$ is

\begin{equation}
\mathrm{i}\omega\epsilon_r\vec{E}_r=\frac{\mathrm{i} m}{r}\vec{H}_z-\frac{\partial \vec{H}_\phi}{\partial z}.
\end{equation}

Using the Yee cell \cite{yee1966} in figure 1c for discretization, the FDFD curl equation for $\vec{E}_r$ can be written as
\begin{eqnarray}
\mathrm{i}\omega\epsilon_r\vec{E}_r(n_r+1/2,n_z) & = &
\frac{\mathrm{i} m}{(n_r+1/2)\Delta r}\vec{H}_z(n_r+1/2,n_z) - \nonumber \\
& & \frac{1}{\Delta z}\left[\vec{H}_\phi(n_r+1/2,n_z+1/2) -
\vec{H}_\phi(n_r+1/2,n_z-1/2)\right].
\end{eqnarray}
The other equations can be discretised similarly.

\section{Hybridizing the T-matrix method with FDFD}
The T-matrix is an operator ($\mathbf{T}$) which acts on the coefficients of the incoming field to produce the coefficients of the outgoing field
\begin{equation}
\vec{p} = \mathbf{T}\vec{a}
\end{equation}
where $\vec{a}$ represents the vector made up of the coefficients ($a_{nm}$ and $b_{nm}$) of the incoming field and $\vec{p}$ represents the vector of the coefficients ($p_{nm}$ and $q_{nm}$) of the outgoing field. The electric fields (and similarly for magnetic fields) can be expanded in terms of incoming and outgoing Vector Spherical Wave Functions (VSWFs)
\begin{equation}
\vec{E}_\mathrm{in} = \sum^{\infty}_{n=1}\sum^{n}_{m=-n} a_{nm} \vec{M}^{(2)}_{nm}(k_{out}\mathbf{r}) + b_{nm} \vec{N}^{(2)}_{nm}(k_{out}\mathbf{r}),
\end{equation}
\begin{equation}
\vec{E}_\mathrm{out} = \sum^{\infty}_{n=1}\sum^{n}_{m=-n} p_{nm} \vec{M}^{(1)}_{nm}(k_{out}\mathbf{r}) + q_{nm} \vec{N}^{(1)}_{nm}(k_{out}\mathbf{r}).
\end{equation}
where $k_{out}$ is the wave vector outside the particle, and $\vec{M}$ and $\vec{N}$ are vector spherical wave functions (VSWFs) defined in \cite{mish1991}. Naturally, we cannot take the sums to infinity but rather taken to $N_{max}$ which is based on criteria defined in \cite{niem2003}. In our model, we would have a dielectric region within the computational grid that would interact with the incoming and outgoing fields. So, in coupling the electric field $\vec{E}(r)$ from the FDFD solutions with the VSWFs for the TE incident modes we obtain
\begin{equation}
\vec{M}^{(2)}_{n'm'}(r) + \sum^{\infty}_{n=1} p_{nm}\vec{M}^{(1)}_{nm}(r) + q_{nm}\vec{N}^{(1)}_{nm}(r) = \vec{E}(r),
\end{equation}
where $n'$ is the incident mode. Similarly for the TM modes,
\begin{equation}
\vec{N}^{(2)}_{n'm'}(r) + \sum^{\infty}_{n=1} p_{nm}\vec{N}^{(1)}_{nm}(r) + q_{nm}\vec{M}^{(1)}_{nm}(r) = \vec{E}(r).
\end{equation}
Due to the rotational symmetry, there is no coupling to other azimuthal modes (i.e. only one value of $m'$ appears). Therefore, all fields share an azimuthal dependence of $\exp(\mathrm{i} m\phi)$.
\begin{figure}[h!]
\begin{center}
{a)\includegraphics[height=5.4cm]{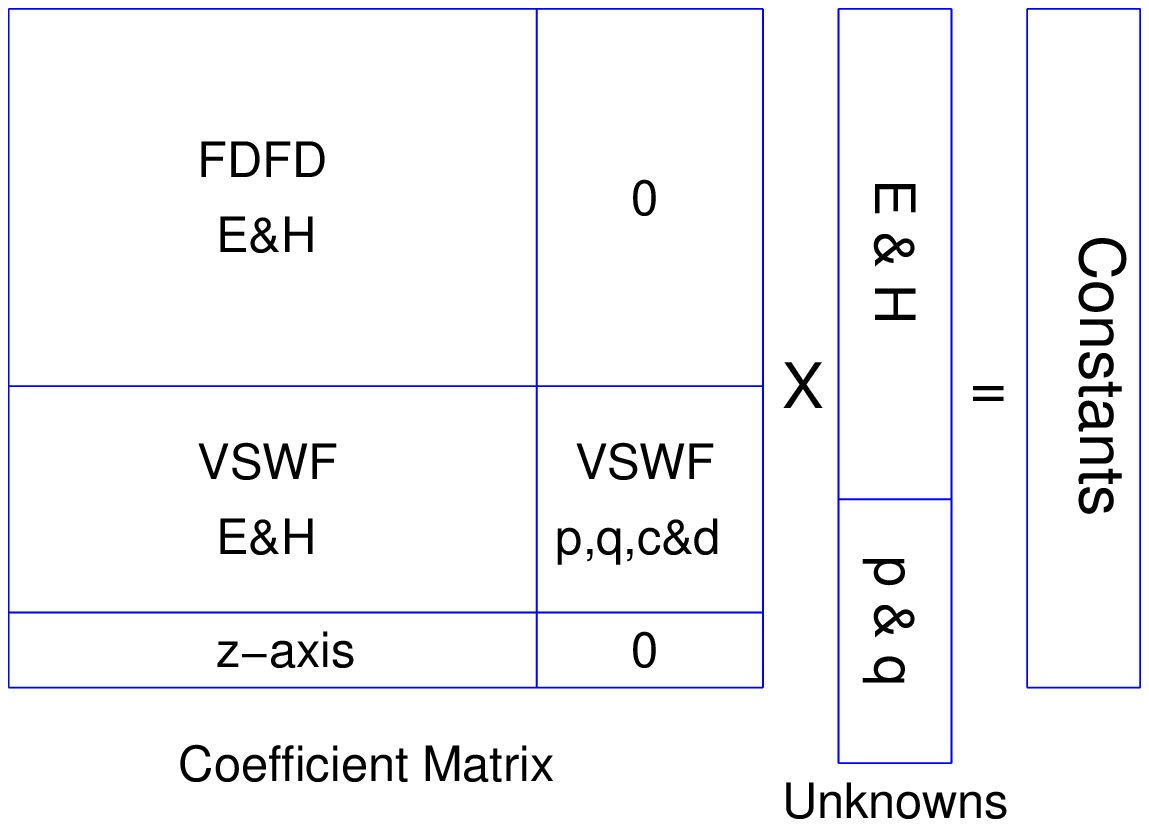} b)\includegraphics[height=5.9cm]{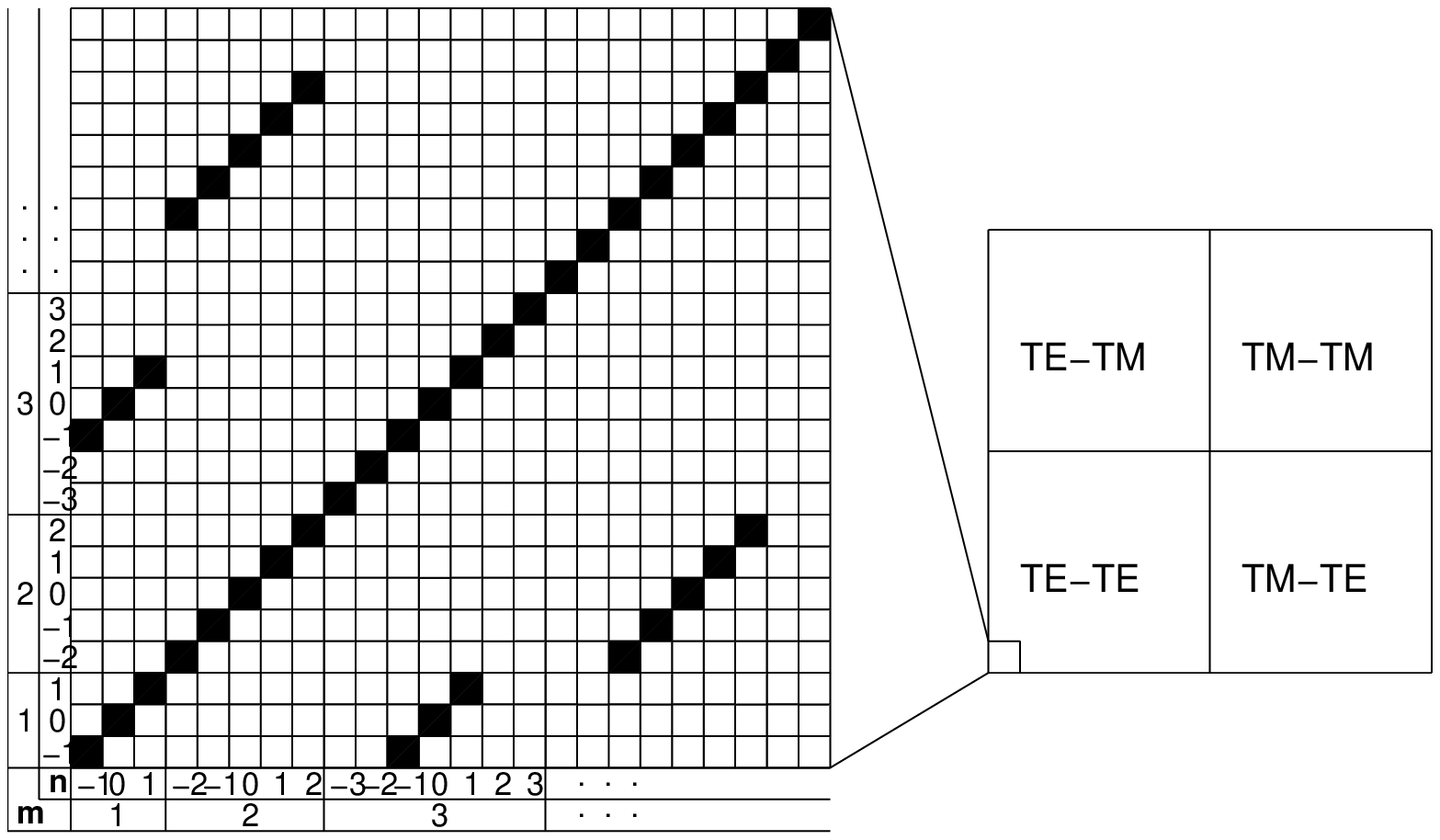}}
\caption{a)\space Coefficient Matrix, Unknowns and Constants. b)\space The T-matrix is made up of four blocks that result from the coupling between the TE and TM modes. The inset is a section of the T-matrix in detail where the allowable coupling between $m$ and $n$ modes are shown as black pixels.}
\end{center}
\end{figure}
Equation (8) or (9) connects the VSWF description of the external fields and the FDFD grid. The VSWF and FDFD equations form an overdetermined linear system (figure 2a) and can be solved using a standard numerical library. The FDFD equations are inserted in the Coefficient matrix (figure 2a) first followed by VSWF equations. Last, the \textit{z}-axis boundary equations are inserted in the Coefficient matrix. Generally, the field is zero at the \textit{z}-axis except for the modes $m=\pm 1$, in which case the first derivatives of the fields are zero. Cycling through all incident modes, the solutions for $p_{nm}$ and $q_{nm}$ are solved given one incident mode at a time and their values are inserted into the T-matrix column representing coupling between the $m$ and $n$ modes (figure 2b).

\section{Discussion}

The micromachines of interest may or may not have $xy$-plane mirror symmetry but they will typically have \textit{n}th-order rotational symmetry. Nonetheless, as with the cube we had modelled, the rotational symmetry can be exploited to reduce the calculation time by orders of magnitude. Conventional T-matrix methods are limited in their application to modelling homogeneous and isotropic materials, with shapes that are close to spheroidal. The FDFD hybridization extends the modelling capability to include \textit{n}th-order rotationally symmetric micro-machines with complex shapes made from materials that are inhomogenous and anisotropic e.g. birefringent crystals.

The Matlab script for solving the matrices in figure 2a was tested on a PC with a 32-bit single 3GHz CPU and 1Gb of RAM. We performed the calculation simulating a 3000\,nm radius cylinder with grid sizes from 1000\,nm--250\,nm. Extrapolating from the natural log scale plot (figure 3), we estimated that it would take 13.6 hours and 165.9 hours (7 days) to perform the calculation given 100\,nm and 50\,nm grid spacing respectively. 

While the foregoing is directed at modelling rotational symmetric particles, we intend to model more complex particles buy using a 3D FDFD grid or the Discrete Dipole Approximation (DDA) method \cite{mish1991} coupling with the VSWFs.
\begin{figure}[!htbp]
\begin{center}
\includegraphics[width=0.7\textwidth]{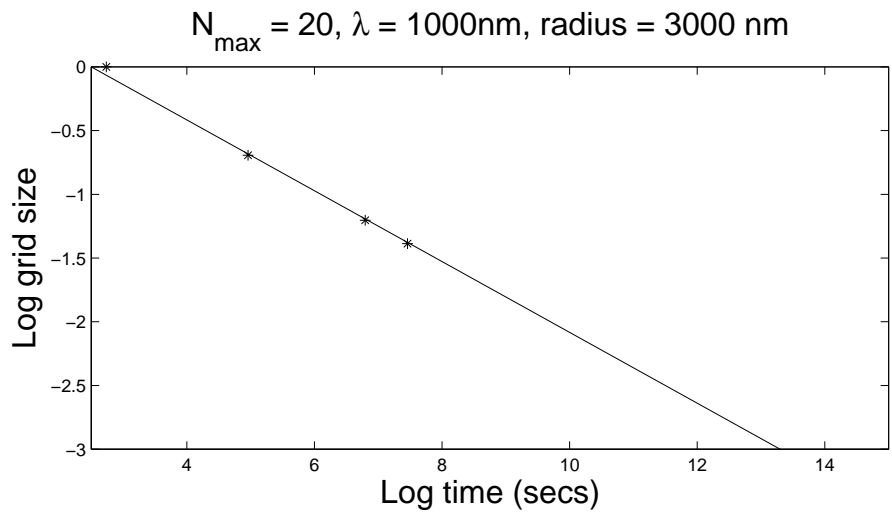}
\caption{Log gridsize in wavelength units versus log time (secs).}
\end{center}
\end{figure}

\end{document}